\documentclass[12pt,preprint]{aastex}

\begin{document}


\title{A $22\arcdeg$ Tidal Tail for Palomar 5}

\author{C. J. Grillmair}
\affil{Spitzer Science Center, 1200 E. California Blvd., Pasadena,  CA 91125}
\email{carl@ipac.caltech.edu}
\and
\author{O. Dionatos\altaffilmark{1}}
\affil{INAF - Osservatorio Astronomico di Roma, via de Frascati 33,
00040, Monteporzio Catone, Italy}

\email{dionatos@mporzio.astro.it}

\altaffiltext{1}{Department of Physics, University of Athens, Pan/polis, 15771 Athens, Greece}

\begin{abstract}

Using Data Release 4 of the Sloan Digital Sky Survey, we have applied
an optimal contrast, matched filter technique to trace the trailing
tidal tail of the globular cluster Palomar 5 to a distance of 18.5
degrees from the center of the cluster. This more than doubles the
total known length of the tail to some 22 degrees on the sky.  Based
on a simple model of the Galaxy, we find that the stream's orientation
on the sky is consistent at the $1.7 \sigma$ level with existing
proper motion measurements. We find that a spherical Galactic halo is
adequate to model the stream over its currently known length, and we
are able to place new constraints on the current space motion
of the cluster.

\end{abstract}



\section{Introduction}

The tidal tails of globular clusters are very interesting from a
dynamical standpoint as they are expected to be very cold
\citep{comb99}. This should make them useful for constraining not only
the orbits of the clusters themselves, but for probing both the global
mass distribution of the Galaxy and its lumpiness
\citep{mura99}. Tidal tails in globular clusters were first discovered
in photographic surveys (\citet{grill95}; \citet{leon2000}) and now
number over 30 in both the Galaxy and in M 31 \citep{grill96}. Most
recently, \citet{gril2006} detected a tidal stream associated with NGC
5466 which subtends some $45\arcdeg$ on the sky.

Among the first discoveries in the Sloan Digital Sky Survey (SDSS)
data were the remarkably strong tidal tails of Palomar 5
\citep{oden2001,rock2002, oden2003}, spanning over $10\arcdeg$ on the
sky. \citet{gril2001} found evidence for a stellar mass function
relatively depleted of lower mass stars, and \citet{koch2004} found
that these lower mass stars have indeed made their way into Pal 5's
tidal stream. Pal 5 may well be on its last orbit around the Galaxy
before dissolving completely.

In this paper we examine Data Release 4 of the SDSS, which encompasses
an area not available to \citet{oden2003}.  We briefly describe our
analysis in Section \ref{analysis}. We discuss our findings in Section
\ref{discussion}, and make concluding remarks in Section
\ref{conclusions}.

\section{Data Analysis \label{analysis}}

Data comprising $u',g',r',i'$, and $z'$ photometric measurements and
their errors for $3.7 \times 10^6$ stars in the region $224\arcdeg <$
R.A. $< 247\arcdeg$ and $-3\arcdeg < \delta < +10\arcdeg$ were
extracted from the SDSS DR4 database using the SDSS CasJobs query
system. The data were analyzed using the matched-filter technique
described by \citet{rock2002}, and the reader is referred to this
paper for a complete description of the method. Briefly, we
constructed an observed color-magnitude density or Hess diagram for
Pal 5 using stars within $5\arcmin$ of the cluster center. An optimum
star count weighting function was created by dividing this
color-magnitude distribution function by a similarly binned
color-magnitude distribution of the field stars. The Hess diagram for
field stars was created using several regions to the east and west of
Pal 5 and the stream, together subtending $\sim 147$ square degrees.
The resulting weighting function was then applied to all stars in the
field and the weighted star counts were summed by location in a two
dimensional array.

We used all stars with $16 < g' < 22$, dereddened as a function of
position on the sky using the DIRBE/IRAS dust maps of
\citet{schleg98}. The distribution of $E(B-V)$ across the field is
shown in Figure 1 and ranges from 0.02 along the northern edge of the
field to 0.35 along the southern edge. 

We optimally filtered the $g' - u'$, $g' - r'$, $g' - i'$, and $g' -
z'$ star counts independently and then co-added the resulting weight
images.  As expected, most of the filter weight is given to main
sequence turn-off and horizontal branch stars, as these populations
lie blueward of the vast majority of field stars. In Figure 2 we show
the co-added, filtered star count distribution for stars on the giant
branch and the main sequence. The image has been smoothed with a
Gaussian kernel of width $\sigma = 0.15\arcdeg$. A low-order,
polynomial surface was fitted and subtracted from the image to remove
large scale gradients due to the Galactic disk and bulge. The blank
area running eastward from just north of Pal 5 and the several blank
areas on the eastern end of the field are regions not included in the
SDSS DR4. We note that the region just north of the cluster
(containing the globular cluster M 5) {\it is} included in the dataset
considered by \citet{oden2003}, and the reader is referred to this
work for a glimpse of Pal 5's trailing tail in the region $1.3\arcdeg
< r < 3.2\arcdeg$ from the cluster.

\section{Discussion \label{discussion}}

Figure 2 clearly shows a long stream of stars with Pal 5's
color-luminosity distribution, extending from R.A., dec =
($226.3\arcdeg, -2.9\arcdeg$) to ($246\arcdeg, +7.9\arcdeg$). The
southern, leading tail has already been described in detail by
\citet{oden2003}, as has the portion of the northern tail westward of
R.A. = $234\arcdeg$. The new result here is the additional $12\arcdeg$
of trailing tidal tail extending to R.A. = $246\arcdeg$. The feature
is quite strong, and appears as a continuous extension of the trailing
tail identified by \citet{oden2003}.  The feature vanishes if the
weighting filter is shifted 0.5 magnitudes blueward or redward of
the Pal 5's observed main sequence.

Comparing Figures 1 and 2, there are clear correlations between
regions of high color excess and areas of reduced star count density.
(e.g. $2\arcdeg$ SE of the cluster, and at R.A., dec = $239\arcdeg,
3.3\arcdeg$). This is to be expected, given that SDSS sample
completeness will be a function of apparent magnitude. The
distribution of foreground dust will affect both the apparent shape
and the number counts along Pal 5's tidal tails. The effect of the
dust can be reduced by counting only the brighter stars, but at some
considerable cost to signal to noise ratio.  This is particularly true at
the eastern end of the tail, where contamination by foreground bulge
stars is almost twice what it is in the vicinity of the cluster
itself.

To investigate whether the additional arc of Pal 5's tail could be due
to confusion with background galaxies we have reanalyzed the same
survey area using as input those objects classified in DR4 as
galaxies. We find no indication of significant enhancements in the
galaxy counts, either extended along the feature, or as discrete
components of it. We conclude that star-galaxy confusion cannot be
held to account for the extended northern feature.

For all these reasons, we conclude that the feature extending from
R.A., dec = ($234\arcdeg, 3.2\arcdeg$) to ($246\arcdeg, 7.9\arcdeg$)
is a {\it bona fide} portion of Pal 5's trailing tidal tail. As found
by \citet{oden2003} for the inner portion of the tails, the linear
density of stars continues to fluctuate along the extended tail,
rising to more than 200 stars per linear degree at R.A. $\approx
240.6\arcdeg$, and dropping almost to indetectablity at R.A. $\approx
236.8\arcdeg$ and $242.8\arcdeg$.  We do not expect significant
diminution in the filter output due to changes in the average distance
between the Sun and the stream. Based on the best-fit orbit model
described below, the distance between us and the stream increases from
23.2 kpc at the cluster itself to 23.9 kpc at apogalacticon (which occurs at
R.A. $\approx 235\arcdeg$), dropping back down to 23.2 kpc at R.A. =
$246\arcdeg$.  Given the photometric uncertainties, this variation in
distance modulus (0.06 mag) is too small to have any noticeable
effect on the output of a matched filter which is optimized for stars
at the distance of Pal 5.  The density fluctuations along the stream
must therefore be real and are presumably a natural consequence of the
episodic nature of tidal stripping of a cluster on an eccentric and
disk-crossing orbit.

In Figure 3 we show the distribution of 78 candidate blue horizontal
branch stars. There is no obvious tendency for these stars to be
distributed along the tidal tails - only four stars outside the
cluster itself could be said to lie along the tails. This is probably
not surprising, however, given that Pal 5 has very few blue HB stars; 
most of the points in Figure 3 are likely to be foreground contaminants.

Assuming that tidal streams closely parallel the orbits of their
parent clusters \citep{oden2003}, we can use the observed orientation
of the extended stream to better constrain the current velocity and
global orbit of Pal 5. Using the analytic Galactic model of
\citet{allen91} (which includes a disk as well as a spherical bulge
and halo), we integrate along the orbit of Pal 5 both forwards and
backwards and project this path onto the sky. We adopt $R_\odot =
8.5$ kpc and $v_c = 220$ km s$^{-1}$, and use \citet{oden2002}'s
heliocentric radial velocity measurement for Pal 5 of $-58.7 \pm 0.2$
km s$^{-1}$

Using Cudworth's 1998 (unpublished, but listed in \citet{dine1999})
proper motion measurement of $\mu_\alpha \cos(\delta), \mu_\delta =
(-2.55 \pm 0.17, -1.93 \pm 0.17$) mas yr$^{-1}$, we arrive at an expected
orbit projection as shown by the solid line in Figure 3.  Other
proper motion measurements for Pal 5 (e.g. \citet{schw1993};
\citet{scho1998}) result in orbit projections which are almost
perpendicular to the stream in Figure 3. With due allowance for
our incomplete knowledge of the Galactic potential and for the
limitations of \citet{allen91}'s model, the projected orbits implied
by the existing proper motion measurements do not agree particularly
well with the orientation of Pal 5's tidal tails.

Based on the width of the ``S'' curvature of the tails in the
immediate vicinity of Pal 5, we adopt a mean offset between Pal 5's
projected orbit and the centerlines of the tails of $0.2\arcdeg$. We
then define a set of normal points which trace the highest surface
densities along the tails in Figure 2. Finally, we project orbits for
a grid of possible proper motions and use $\chi^2$ minimization to
obtain a ``best fit''. The minimum $\chi^2$ solution is shown by the
dotted line in Figure 3. 

Over the length of the stream for which we have good signal to noise
ratio, our best fit orbit parallels the stream contours quite
well. Departures appear to be of high order rather than systematic,
indicating that a spherical halo potential is adequate to fit the
data. There is a suggestion that the southern tail and the model may
be parting company near the limits of the field, and it will be
interesting to see whether the simple potential used here can be made
to fit the continuation of the southern tail when additional data
become available.

The proper motion which corresponds to our best fit orbit has
$\mu_\alpha \cos(\delta), \mu_\delta = (-2.27 \pm 0.08, -2.19 \pm
0.03)$ mas yr$^{-1}$, where the uncertainties correspond to the 99\%
confidence interval. We note that, while the uncertainties are
influenced by measurement errors, random motions of stream stars,
confusion with the foreground population, variable extinction across
the stream, they do not incorporate uncertainties in our adopted model
for the Galaxy. Our modeled proper motions are within $\sim 1.7
\sigma$ of Cudworth's measured values. For the orbit shown in Figure
3, the corresponding space velocities of Pal 5 are U,V,W = $(48.3 \pm
2.0 , -334 \pm 5, -14.2 \pm 2.1)$ km s$^{-1}$. The (radial) period of
the orbit is $2.9 \times 10^8$ yrs, with peri and apogalacticon of 7.9
and 18.8 kpc, respectively. These values agree very well with similar
estimates by \citet{oden2003} based on their less extensive data set.

The best fit orbit predicts a total length of $\simeq 8.3$ kpc for the
northern arm. The width of the tail appears roughly constant along its
length, though the signal to noise ratio is low enough that we cannot
rule out a possible widening of the tail beyond R.A. = $241.6$.
Comparing the contours in Figure 3 with the best fit orbit, there is
also a suggestion of ``wobbling'' of the stream eastwards of R.A. =
$239\arcdeg$, and perhaps a $0.5\arcdeg$ northward jog of the tail at
$241.5\arcdeg < $ R.A. $< 244.5$. If borne out by future kinematic
data, this could be evidence for irregularities in the Galactic
potential or for a weak encounter between stream stars and a
substantial mass concentration (e.g. a large molecular cloud) during a
recent passage through the disk. A more extensive analysis of the
stream and its consequences will be presented in a forthcoming paper.

\section{Conclusions \label{conclusions}}

Applying optimal contrast filtering techniques to SDSS DR4 data, we
have detected a continuation of Pal 5's trailing tidal stream out to
almost $19\arcdeg$ from the cluster. Combining this with the already
known southern tail of Pal 5 yields a stream some $22.5\arcdeg$ long
on the sky. The extension of the stream shows marked fluctuations in
surface density along its length which are presumably a natural
consequence of the episodic nature of tidal stripping.  The
orientation of the stream is in accord (at the $1.7 \sigma$ level)
with one of the existing measurements of Pal 5's proper motion. The
stream can be modeled reasonably well using a model Galactic potential
with a spherical halo, and we are able to place much better
constraints on the current space motion of the cluster.

We can use the current data set to assign to each star a probability
that it is associated with Pal 5's tidal stream. These probabilities
can be taken to the telescope, where we will need to measure radial
velocities sufficiently accurately to unambiguously tie the stars to the
stream.  Ultimately, the vetted stream stars will become prime targets
for the Space Interferometry Mission, whose proper motion measurements
will enable very much stronger constraints to be placed on both the
orbit of the cluster and on the potential field of the Galaxy.

\acknowledgments

We are grateful to an anonymous referee for observations which improved the clarity of the manuscript. 

Funding for the creation and distribution of the SDSS Archive has been
provided by the Alfred P. Sloan Foundation, the Participating
Institutions, the National Aeronautics and Space Administration, the
National Science Foundation, the U.S. Department of Energy, the
Japanese Monbukagakusho, and the Max Planck Society.

{\it Facilities:} \facility{Sloan}.

\clearpage



\begin{figure}
\epsscale{1.0}
\plotone{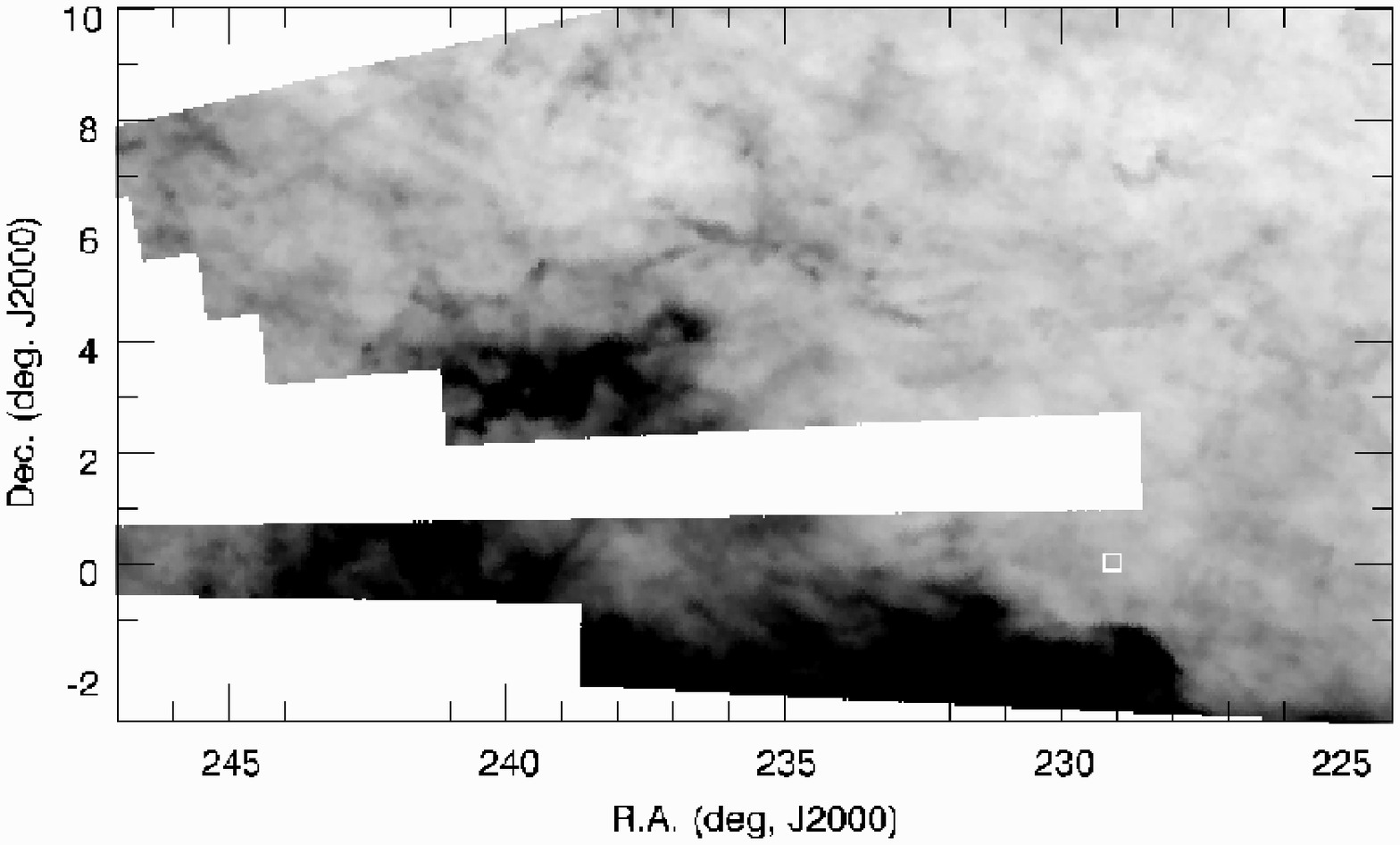}
\caption{Distribution of E(B-V) over the DR4 field surrounding Pal
5. The position of Pal 5 is indicated by the open square at R.A., dec
= (229,-0.11). The color excess ranges from 0.02 along the northern
border of the image, to 0.3 along the southern edge. \label{fig1}}
\end{figure}

\begin{figure}
\epsscale{1.0}
\plotone{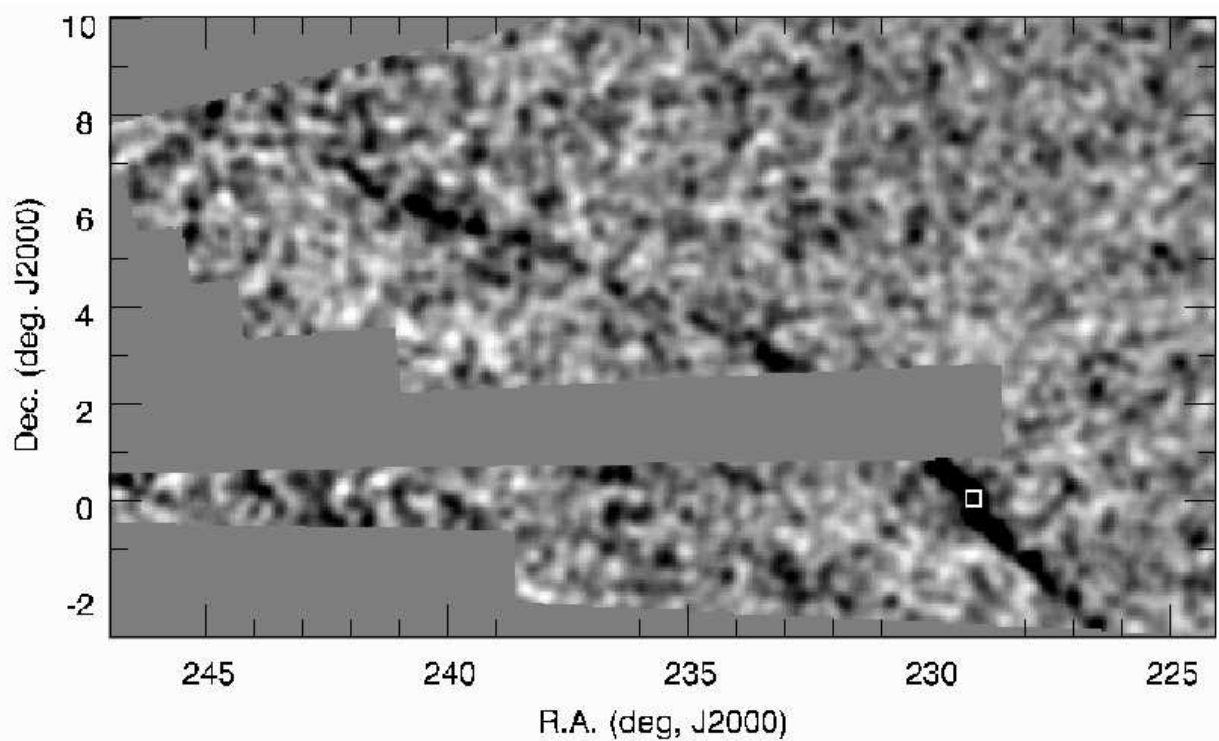}
\caption{Smoothed, summed weight image of the SDSS field after
subtraction of a low-order surface fit. Darker areas indicate higher
surface densities. The location of Pal 5 is indicated by the open
square at R.A., dec = (229,-0.11). The weight image has been smoothed
with a Gaussian kernel of width of $0.15\arcdeg$. The irregular
borders and the missing stripe are defined by the limits of SDSS Data
Release 4.  The newly discovered extension of Pal 5's tidal stream
extends from R.A., dec = ($234\arcdeg, 3.2\arcdeg$) to ($246\arcdeg,
7.9\arcdeg$). \label{fig2}}
\end{figure}

\begin{figure}
\epsscale{1.0}
\plotone{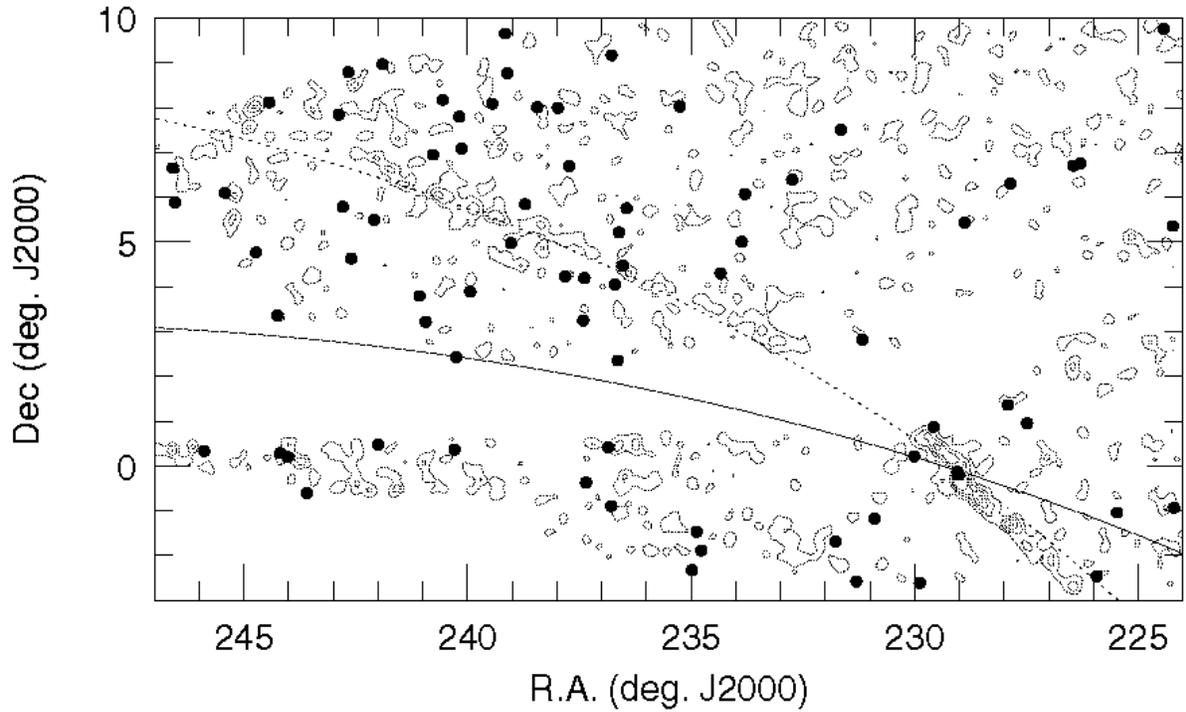}
\caption{Surface density contours and the orbit of Pal 5. Contours are
taken from the weight image in Figure 2 and represent the 2, 3, 4, and
$5 \sigma$ levels. The filled circles show the positions of candidate
blue horizontal branch stars. The solid line shows an orbit
integration based on the Cudworth's 1998 measurement of the cluster's
proper motion \citep{dine1999}. The dotted line shows an orbit with
$\mu_\alpha \cos(\delta), \mu_\delta = (-2.27, -2.19)$ mas yr$^{-1}$.}
\end{figure}

\end{document}